\documentclass[5p]{elsarticle}

\usepackage{siunitx}
\usepackage{amsmath}
\usepackage{color}
\usepackage[list=off]{caption}
\usepackage{xspace}
\usepackage{verbatim}
\usepackage{hyperref}
\usepackage{natbib}

\renewcommand\Re{\operatorname{Re}}
\renewcommand\Im{\operatorname{Im}}

\newcommand{\blue}{}

\newcommand{\bitfigscal}{0.4}

\newcommand{\bshuf}{{\tt Bit\-shuffle}}
\newcommand{\shuf}{{\tt shuffle}}
\newcommand{\deflate}{{\tt DEFLATE}}
\newcommand{\lzss}{{\tt LZ77}}
\newcommand{\lzfour}{{\tt LZ4}}
\newcommand{\lzfourhc}{{\tt LZ4\_HC}}
\newcommand{\lzf}{{\tt LZF}}
\newcommand{\blosc}{{\tt Blosc}}

\renewcommand{\a}{\alpha}
\renewcommand{\b}{\beta}

\biboptions{square,numbers,compress}

\begin{document}

\title{A compression scheme for radio data in high performance computing}


\author[ubc,cifar]{Kiyoshi~Masui\corref{cor0}}
\author[ubc]{Mandana~Amiri}
\author[cita,dunlap,ut]{Liam~Connor}
\author[ubc]{Meiling~Deng}
\author[ubc]{Mateus~Fandino}
\author[ubc]{Carolin~H\"ofer}
\author[ubc]{Mark~Halpern}
\author[mcg]{David~Hanna}
\author[ubc]{Adam~D.~Hincks}
\author[ubc]{Gary~Hinshaw}
\author[mcg]{Juan~Mena~Parra}
\author[dunlap]{Laura~B.~Newburgh}
\author[ubc,cita]{J.~Richard~Shaw}
\author[ut,dunlap]{Keith~Vanderlinde}

\address[ubc]{Department of Physics and Astronomy, University of British Columbia, 6224 Agricultural Rd. Vancouver, V6T 1Z1, Canada}
\address[cifar]{Canadian Institute for Advanced Research, CIFAR Program in Cosmology and Gravity, Toronto, ON, M5G 1Z8}
\address[cita]{Canadian Institute for Theoretical Astrophysics, 60 St George St, Toronto, ON, M5S 3H8, Canada}
\address[dunlap]{Dunlap Institute for Astronomy \& Astrophysics, University of Toronto, 50 St George St, Toronto, ON, M5S 3H4, Canada}
\address[ut]{Department of Astronomy \& Astrophysics, University of Toronto, 50 St George St, Toronto, ON, M5S 3H4, Canada}
\address[mcg]{Department of Physics, McGill University, 3600 University St, Montreal, Canada}

\cortext[cor0]{Corresponding author, kiyo@physics.ubc.ca}


\date{\today}

\begin{abstract}

    We present a procedure for efficiently compressing astronomical radio data for
    high performance applications. Integrated, post-correlation data are first
    passed through a nearly
    lossless rounding step which compares the precision of the data to a
    generalized and calibration-independent form of the radiometer equation.
    This allows the precision of the data to be reduced in a way that has an
    insignificant impact on the data.  The newly
    developed \bshuf{} lossless compression algorithm is subsequently applied.
    When the algorithm is used in conjunction with the HDF5 library and data format,
    data produced by the CHIME Pathfinder telescope is compressed 
    to 28\% of its original size and decompression throughputs
    in excess of 1\,GB/s are obtained on a single core.

\end{abstract}


\maketitle

\section{Introduction}


The simultaneous drives to wider fields and higher sensitivity have led radio
astronomy to the cusp of a big-data revolution. There is a multitude
of instruments,
including 21\,cm cosmology experiments 
\citep{
    BAOBAB,
    BINGO,
    CHIME,
    Pober2013,
    LEDA,
    LOFAR,
    MITEoR,
    PAPER,
    TIANLAI},
Square Kilometer Array Precursors
\citep{
    ASKAP,
    MWA,
    MeerKAT},
and ultimately the Square Kilometer Array \citep{SKA}, whose rate of data
production will be orders of magnitude higher than any existing radio 
telescope. An early example is the CHIME 
Pathfinder \citep{2014SPIE.9145E..22B, 2014SPIE.9145E..4VN} which will soon be
producing data at a steady rate of over 4\,TB per day. The
cost associated with storing and handling these data can be considerable and
therefore it is desirable to reduce the size of the data as much as possible using
compression. At the same time, these data volumes produce a significant data
processing challenge. Any data compression/decompression scheme must be fast
enough as to not hinder data processing, and would ideally lead to a net
increase in performance due to the reduced time required to read the data from
disk.

Here, after discussing some general considerations for designing data storage
formats in Section~\ref{s:cons}, we present a scheme for compressing astronomical 
radio data.  Our
procedure has two steps: a controlled (relative to thermal noise) reduction of
the precision of the data which reduces its information entropy
(Section~\ref{s:lossy}), and a lossless
compression
algorithm---\bshuf{}\footnote{\url{https://github.com/kiyo-masui/bitshuffle}}---which
exploits this reduction in entropy to achieve a
very high compression ratio (Section~\ref{s:lossless}).
These two steps are independent in that, while
they work very well together, either of them can be used without the other.
When we evaluate our method in Section~\ref{s:eval} we show that the precision
reduction improves compression ratios for most
lossless compressors.  Likewise, \bshuf{} outperforms most other lossless
compressors even in the absence of precision reduction.

\section{Considerations for designing data storage formats}
\label{s:cons}

\subsection{Characteristics of radio-astronomy data and usage patterns}

Integrated, post-correlation radio-astronomy data are typically at 
least three dimensional, containing
axes representing spectral frequency, correlation product, and 
time\footnote{A fourth axis is
  often introduced when data are `folded' or `gated'---i.e., if data from the on-
  and off-periods of a switched, calibration noise source are
  accumulated separately, or pulsar data is folded on the pulsar's period which is
  divided into many gates.}.
The \emph{correlation product} refers to the correlation of all 
antenna input pairs, including auto-correlations and cross-correlations between 
different polarisations from the same antenna. In a single dish these
form the polarization channels for each beam and in an interferometer these are
the visibilities. This also applies to beam forming interferometers, where
linear combinations of antenna inputs are formed (either in analog or
digitally) before correlation.

The CHIME collaboration determined that its data are most commonly
accessed along the time axis. That is, it is generally most efficient
for the axis representing time to be the fastest varying once loaded
into memory. This is the case for noise characterisation, radio-frequency
interference (RFI)
flagging and system-health monitoring, to name a few. Most
importantly, the map-making pipeline typically produces maps on a
per-frequency basis and is most efficient at processing
time-contiguous data. Though it is sometimes necessary to work with
spectra (slices along the frequency axis) or `correlation triangles'
(slices along the correlation product axis), we find that these use cases normally
only involve a few slices and large I/O operations in these spaces are rare.

Of course, the CHIME collaboration's preference for the time axis to be the fastest
varying will not apply to all consumers of radio data. \blue{One expects that
access patterns might vary considerably for the diverse applications of radio
data, including spectroscopy, synthesis imaging, and pulsar timing.}
But as discussed below, arranging data with
time as the fastest varying index is beneficial for data compression.

\subsection{Compression: benefits and requirements}
\label{s:req}

Compression can greatly ease the burden of storing and handling large data
sets, but there are also
performance benefits. Compression algorithms exist whose
decompression cost is negligible compared to the cost of reading from
disk. As we will show, data may be compressed by up to a
factor of four in some cases. As such, the time required to load a
dataset from disk into memory may be reduced by a factor of four using
compression.

We previously stated that ordering data with the axis representing time as the
fastest varying is most efficient for the majority of I/O operations. This
ordering is also beneficial for compression, since adjacent data
points are likely to be highly correlated, 
presuming that the cadence is such that the
spatially-smoothly varying sky is Nyquist sampled.
On the other hand, it is most natural to record data with time as the
\textit{slowest} varying index since that is the order in which they are
generated by the
instrument. To have time as the fastest varying index,
the data must either be buffered in memory (which is impractical),
written with strided writes (which is inefficient) or reordered after
acquisition. Since the data are acquired and written only once but
read many times, it is logical to prioritise read-performance over
write-performance. Thus, the CHIME collaboration deemed a
post-acquisition reordering step to be worthwhile. 

The same argument can be used to prioritise data decompression speed
over compression speed. Compression is sufficiently cheap computationally that
even a modest number of processors should be able to keep up with the
acquisition rate of CHIME Pathfinder data (which will be 
$\sim50\,{\rm MiB/s}$ depending on runtime parameters)
for almost any compression algorithm. Even if this were not
the case, data could be compressed in parallel post acquisition.  On the other
hand, one might wish to load several days of acquired data at once for
analysis, and ideally, this would be bound only by disk read times, not
decompression speed.

The decompression cost may not be negligible compared to read times
for files that are cached in memory or stored on high-performance
parallel file systems. This makes it desirable to have as fast a
decompression scheme as possible as the benefits of speed
are not always limited by hard drive access. A multi-threaded implementation of
the decompression algorithm can thereby result in a significant speed up on
multi-core systems.

To summarise all of the foregoing, the following requirements for a
compression scheme emerge:

\begin{description}
    \item[Unbiased] Any lossy compression must not bias the data in any way.
    \item[Nearly lossless] Any lossy compression employed must be controlled in
        a manner that is \emph{guaranteed} not to significantly decrease the
        sensitivity of the data.
    \item[Time minor] In the multi-dimensional dataset, the axis representing
        time should be the fastest varying.  This allows for the efficient
        reading of small subsets of spectral frequencies and correlation
        products but for large periods of time.
    \item[Fast decompression] To realize the performance gains associated with
        compressing the data, we require the time to decompress the data to be
        small compared to the time required to read the data from disk. 
        At the time of writing, a single hard
        drive can typically be read at a rate of $\sim100\,{\rm MiB/s}$. As
        such, a compression algorithm with throughput of $\sim 1\,{\rm GiB/s}$
        on a single processor is desirable.
    \item[Threaded] When using a parallel file system, or
        when the file is cached in system memory, reading throughputs can be
        much higher compared to when using a single hard disk. For
        decompression to not degrade performance in these cases, the compression
        library should be threaded.
    \item[Thread-safe] While the HDF5 library (see
        Section~\ref{ss:hdf5}, below) is not internally threaded, it
        may become so in the future.  In addition, programs
        may attempt to hide the cost of IO operations by putting them in a
        separate thread.  The compression library must therefore be
        thread-safe.
\end{description}

\subsection{HDF5 and chunked storage}
\label{ss:hdf5}

The Hierarchical Data Format 5 (HDF5) \citep{hdf5} is a widely used
data format in astronomy, capable of storing and organizing large
amounts of numerical data. In the context of this paper, it also has
the benefit of allowing for `chunked storage'. That is, an
HDF5 `dataset'---a multidimensional array of homogeneous type---can be
broken up into subsets of fixed size, called chunks, which are stored
in any order on disk, with locations recorded in a look-up table. This
is in contrast to contiguous storage, in which random access is
obviously trivial. The advantage of chunked storage for our purposes
is that though the number of elements in a chunk is fixed, its size is
not, and as such it may readily be compressed. 

The primary drawback of chunked storage is that full chunks must be
read from disk at a time.  As such, to read a single array element,
the full chunk containing thousands of elements must be read.  In
practice, this is mitigated by the fact that hard
disk latencies are very long compared to the time required to read
data from disk. Typically a chunk of several hundred KiB can be read
from disk in only twice the time required to read a single element of
a contiguous dataset, and thus the cost of chunked storage for random
access is at most a factor of two, as long as chunk sizes are chosen
appropriately. In addition, requiring a large number of random accesses
to a dataset is a rare usage case \blue{for radio data,}
and therefore random access performance
should rarely be a driving factor when designing a data format.

HDF5 implements compression of chunked datasets via a `filter
pipeline'. Any number of filters can be specified when the dataset is
created. When a data chunk is to be stored, the buffer is sequentially
passed through filter functions before being written to disk.  When the
chunk is to be read from disk, it is passed through the inverse filter
functions in reverse order before being presented to the user. Many
HDF5 filters are available whose functions include lossless and lossy
compression, preconditioning filters aimed at improving compression
ratios, and other utilities such as data check-sums. The
filter-pipeline is also extensible, and writing a new filter, such as
the one presented in this paper, is a relatively straightforward task.

\section{Lossy entropy reduction: reduction of precision}
\label{s:lossy}

All experiments must perform some amount of lossy compression simply by virtue
of having to choose a finite width data type which reduces precision by
truncation.
Here, we focus on performing a reduction of precision in a manner that is
both controlled, in that it has a well-understood effect on the
data; and efficient, in that only the required precision is kept allowing for better
compression.

Reducing the precision of the data involves discarding some number of the
least significant
bits of each data element whose significance is small compared to the noise.
For integers, this is accomplished by rounding values to a multiple of a
power of two.
\blue{For floating point numbers the equivalent operation is performed on the
significand part of the value, however in this work we focus on integer
data.}
In some cases this may allow for data to fit into a smaller data type, for
example single-precision (32-bit) as opposed to double-precision (64-bit)
floating point numbers.  However, even if this is not the case it is still
beneficial to identify bits that are well within the noise margin and replace
them with zeros. Bits that are dominated by noise are essentially
random and are thus very high entropy.  Any subsequent lossless compression has
no hope of compressing them despite their insignificance.

Of course for this to be useful, it is necessary for the lossless
compression step
to be able to exploit the reduced entropy associated with zeroing the noisy
bits.  This will be discussed in Section~\ref{s:lossless}.

\blue{
In this section, we begin with a discussion of noise for a general radio
dataset.  We derive an expression for the thermal noise, including possible
correlated components from the sky (so-called self-noise), that is independent
of calibration.  We then use this estimate to derive an acceptable level of
rounding, specified by the rounding granularity, such that the induced error
and thus rounding noise is negligible compared to thermal.  The final result is
a procedure for rounding the data elements that maximizes the reduction of
entropy of the data while constrained to fixed loss of sensitivity.
}

\subsection{Noise and the radiometer equation}

Any discussion of reducing numerical precision must necessarily include a
discussion of noise, the intrinsic scatter in the data independent of any
discretization effects. If the error induced by reducing the precision of the
data is small compared to the scatter from the noise, then the reduction will
have a negligible effect on the data, assuming the error is unbiased.
Here, we will focus on thermal noise
which is present for all sources of incoherent radiation and is thus a lower
limit on the noise present in the data. Thermal noise causes uncertainty in the
measurement of radiation power that is proportional to that power. It is due to
the stochastic nature of incoherent radiation. Coherent sources of radiation
can increase the measured power
without increasing scatter, however such sources are rare in astronomy.
Radio-frequency interference, however, may be coherent.

Fortunately the thermal noise can be estimated from the data on a sample-by-sample
basis, and independently of any calibration factors, using the radiometer
equation. Usually it is assumed that the thermal noise is uncorrelated between
correlation products, but we will show that the noise is in general
correlated and will compute the associated covariance matrix.
We will argue that in some
observational regimes it may be necessary to take these correlations into
account when reducing the numerical precision of the data relative to the
noise.

We denote the correlation products in a single spectral bin and in a single
time integration as $V_{ij} \equiv \langle a_i a_j^* \rangle$, where $a_i$ is
the digitized and Fourier-transformed signal from antenna channel $i$, and the
angular brackets are an ensemble average which is approximated using a time
average within the integration time.  The noise is characterized by the covariance
matrix of the full set of correlation products,
$C_{\a ij,\b gh}$. Here, the indices $i$,
$j$, $g$, and $h$ run over the antenna number and $\a$ and $\b$ run over the real
and imaginary parts.  For example $C_{\Re ij,\Im gh} \equiv \langle \Re V_{ij}
\Im V_{gh} \rangle - \langle \Re V_{ij} \rangle \langle \Im V_{gh} \rangle$.

It is typically assumed that the cross-correlations between channels are much
smaller than the auto-correlations, i.e. $V_{ij} V_{ij}^* \ll V_{ii} V_{jj}$
for $i \ne j$.  This is because it is assumed that the total measured power is
dominated by noise from the amplifiers in the signal chain prior to
digitization.  In this limit, all correlation products are uncorrelated and
the well known radiometer equations are 
\blue{\citep[chap. 6.2]{2001isra.book.....T}}
\begin{align}
    \sigma_{\Re ii}^2 &= \frac{V_{ii}^2}{N}, \\
    \sigma_{\Im ii}^2 &= 0, \\
    \sigma_{\a ij}^2 &= \frac{V_{ii} V_{jj}}{2N} \qquad (i \neq j).
\end{align}
Here $N = \Delta_\nu \Delta_t$ is the number of samples entering the
integration, and one notes that the auto-correlations, $V_{ii}$, are purely
real. These can be aggregated into a diagonal covariance matrix:
\begin{equation}
    \label{e:rad_simple}
    C_{\a ij,\b gh} = \delta_{\a\b}
    (\delta_{ig} \delta_{jh} + \delta_{ih} \delta_{jg})
    (1 - \delta_{ij}\delta_{\a \Im})
    \frac{V_{ii} V_{jj}}{2N},
\end{equation}
\blue{where $\delta_{ij}$ is the Kronecker delta which is unity for $i = j$ and zero
otherwise.} The first factor in parentheses just cancels the factor of two for
auto-correlations and accounts for the fact that $V_{ij} = V_{ji}^*$.
The second factor in parentheses sets the whole expression to
zero for the imaginary part of the auto-correlations.

Equation~\ref{e:rad_simple} is appropriate in many observational regimes and is
the equation we use for CHIME data, However, there are cases when the
correlation products can be highly correlated. With the recent increased 
emphasis on low spectral frequencies (where sky
brightness typically dominates over amplifier noise) and on close packed
arrays (where amplifier noise may become coupled between channels) the
approximation that the cross-correlations are small may not hold.  A more
general set of equations describing the noise is \citep{1989AJ.....98.1112K}:
\begin{align}
    \label{e:rad_full1}
    C_{\Re ij,\Re gh} 
    & = \frac{1}{2N} \Re\left[ V_{ig} V_{jh}^* + V_{ih} V_{jg}^* \right],\\
    \label{e:rad_full2}
    C_{\Im ij,\Im gh} 
    & = \frac{1}{2N} \Re\left[ V_{ig} V_{jh}^* - V_{ih} V_{jg}^* \right],\\
    \label{e:rad_full3}
    C_{\Re ij,\Im gh} 
    & = \frac{1}{2N} \Im\left[- V_{ig} V_{jh}^* + V_{ih} V_{jg}^* \right].
\end{align}
These may be derived
from first principles by Wick expanding the four-point correlations of $a_i$ in
terms of the correlation products $V_{ij}$.

Below we show some special cases of the above equations to illustrate how they
differ from Equation~\ref{e:rad_simple}.
\begin{align}
    C_{\Re ii,\Re ii}
    &= \frac{1}{N} V_{ii}^2\\
    C_{\Im ii,\Im ii}
    &= 0\\
    C_{\Re ii,\Re ij}
    &= \frac{1}{N} V_{ii} \Re (V_{ij})\\
    C_{\Re ij,\Re ij}
    &= \frac{1}{2N} \left[ V_{ii} V_{jj} + \Re(V_{ij})^2 - \Im(V_{ij})^2
    \right]\\
    C_{\Im ij,\Im ij}
    &= \frac{1}{2N} \left[ V_{ii} V_{jj} - \Re(V_{ij})^2 + \Im(V_{ij})^2
    \right]\\
    C_{\Re ij,\Im ij}
    &= \frac{1}{N} \left[ \Re V_{ij} \Im V_{ij} \right]
\end{align}

One would expect that it is the diagonal of this covariance matrix,
$C_{\a ij,\a ij}$, to which the error associated
with precision reduction should be compared. These diagonal elements give the
variance of a data element marginalized over all other elements.
It is actually more appropriate to compare the error to the
unmarginalized variance, that is $1/(C^{-1})_{\a ij,\a ij}$.
If the truncation error is small compared to the
marginalized variance then the ability to measure an individual visibility will
be unaffected, while if the error is small compared to the unmarginalized
variance, then the ability to measure any linear combination of correlation
products
will be unaffected. The two expressions are equivalent
if using Equation~\ref{e:rad_simple} for the covariance but may differ
significantly when the data are dominated by the sky instead of receiver-noise.

As an illustrative example of how the noise can be highly correlated between
correlation products, consider the case of an interferometer
where the visibilities are
dominated by a single unresolved point source.  For
simplicity, we assume the source is at zenith and that the gains and phases
of all inputs are calibrated, although our argument does not depend on this.
For these conditions, the real parts of all visibilities are equal and
proportional to the source flux, while the imaginary parts are zero.
\blue{As shown in \citet{1989AJ.....98.1112K}, the noise is dominated by the
so-called self-noise.}
All the
elements of the $(\Re \Re)$ block of the covariance matrix are also equal, 
indicating that the visibilities are perfectly correlated and
that there is only one effective measurement of the source. However, the
variance of the difference between any two visibilities is zero.  The
difference of the two visibilities is essentially a new correlation product
whose effective beam has a null at the zenith.  So while the
auto-correlations and thus the variance of the visibilities are dominated by
the source, any linear combination of visibilities whose effective beam does
not include the source will have much lower
noise.

When defining the matrix $\bf C$, it is necessary to note that there are
several redundant combinations of indices for the correlation products.
The $(\Re ij)$ index is equivalent
to the $(\Re ji)$ element, and the $(\Im ij)$ and $(\Im ji)$ correlation
products are
related by a negative sign.  In addition the $(\Im ii)$ correlation products are
identically zero, carry no information, and have no noise. The rows and columns
of $\bf C$ associated with these correlation products should be discarded.  However,
after removing this redundancy, $\bf C$ is guaranteed to be at least positive
semi-definite and, as long as the system temperature is finite, positive
definite and thus invertible. Only non-linearity in the \emph{correlator} (not
the analog systems or ADCs) can render it non-invertible.

Unfortunately, for interferometers with more than a few dozen elements, it is
not feasible to invert the covariance matrix for every spectral bin and every
temporal integration in real time.  As such, for large interferometers, we must fall
back to using Equation~\ref{e:rad_simple} over
Equations~\ref{e:rad_full1}--\ref{e:rad_full3}. While not strictly accurate,
Equation~\ref{e:rad_simple} remains an excellent approximation for most
observational regimes. For CHIME, no source on
the sky other than the sun increases the total power by more than roughly
40\% at any spectral frequency. Since the errors from precision reduction will
be sub-dominant to the noise by several orders of magnitude, even order unity
mis-estimations of the noise should have negligible impact on the final data.
Nonetheless, one may want to compensate for this error by being
extra conservative when specifying
the degree of precision reduction as will be discussed in the next section.

For the remainder of this paper we will define the RMS noise used to calculate
truncation precision as
\begin{equation}
    \label{e:noise}
    s_{\a ij} \equiv \sqrt{1/(C^{-1})_{\a ij,\a ij}},
\end{equation}
with $\bf C$
defined in either Equation~\ref{e:rad_simple} or
Equations~\ref{e:rad_full1}--\ref{e:rad_full3}
depending on interferometer size.

\subsection{Rounding the data relative to the noise}

With an expression in hand for $s_{\a ij}$, which is our basis for
comparison when adding numerical noise, we can proceed to derive a procedure
for rounding the data values.
In this section we will drop the indices on
$s$ with the procedure being understood to apply on a correlation product by
correlation product basis.

We will be relating $s$ to the rounding granularity
$g$, which is the power of two to a multiple of which we will be rounding each
data element.
Note that by rounding instead of simply
truncating, one extra bit can be discarded for
essentially no cost since the induced error associated with zeroing a given
number of bits is half as large.
Rounding also has the very desirable property of being unbiased\footnote{We
choose the `round ties to even' tie breaking scheme \citep{ieee754},
which is unbiased.}, provided
the data values are randomly distributed within the granularity (which is an
excellent approximation since the granularity will be much smaller than the Gaussian
thermal noise).  This is not true of truncation, where the bias is half
the granularity.

We define $\sigma_r^2$ as the added noise variance associated with reducing the
precision of a data element. We will require that $\sigma_r^2 < fs^2$, where
$f$ is the maximum fractional
increase in noise from precision reduction. It can be thought of as the effective
fractional loss of integration time caused by reducing the precision.

When using the approximate Equation~\ref{e:rad_simple} in the definition of $s$,
the approximation can be compensated for by reducing $f$ by a factor of the
minimum portion of total power originating from receiver-noise squared. To be
more precise, multiplying $f$ by the minimum of
$1 - V_{ij}V_{ij}^* / (V_{ii} V_{jj})$ (for $i \neq j$),
roughly compensates for the approximation.
\blue{This will be especially relevant to
low-frequency compact arrays with many elements, for reasons mentioned above.}
This could in principle be performed dynamically as a function of time or spectral
frequency, although no attempt is made to implement this.

For randomly distributed rounding errors (which is an excellent approximation
since the rounding will occur in noise-dominated bits), the rounding noise is
related to the granularity by \citep{lathi1998modern}:
\begin{equation}
    \sigma_r^2 = g^2 / 12.
\end{equation}
Thus the \emph{maximum rounding granularity} is:
\begin{equation}
    \label{e:trunc_error}
    g < \sqrt{12fs^2},
\end{equation}
Our precision-reduction scheme is to round each data element to a
multiple of the largest possible power of two, $g$, subject to the constraint 
given in Equation~\ref{e:trunc_error}. \blue{We note that this equation gives
an upper limit, and that on average the granularity and added noise will be
below this limit.}

\blue{
For maximum generality, the set of $s_{\a ij}$ should be recalculated for each
spectral frequency and each temporal integration, allowing the precision
reduction to adapt to sky and bandpass spectral
structure, temporal changes in the sky, and time and frequency dependant RFI.
}

\blue{
When using
Equation~\ref{e:rad_simple} for the noise, the calculation of $s_{\a ij}$
requires only a handful of floating point operations per data-element and thus
has negligible cost compared to the initial correlation.
Finding the largest integer power-of-two granularity that satisfies
Equation~\ref{e:trunc_error} and then rounding to that granularity can be
performed in tens of instructions with no branching.
An example implementation in the Cython programming language is available
online\footnote{\url{https://gist.github.com/kiyo-masui/b61c7fa4f11fca453bdd}}.
}

\blue{
As discussed above, performance is a greater concern for decompression than
compression. The precision reduction requires no decoding, and as such
its throughput is of secondary concern. The example implementation
is only lightly optimized and achieves $\sim300\,{\rm MiB/s}$ throughput on a
single core of a modern processor. This was deemed sufficient for CHIME's data
acquisition, although we hypothesize that a factor of four speed-up may be
possible by employing the vectorized SSE instruction sets.
}

\begin{figure}
    \centerline{\includegraphics[scale=0.4]{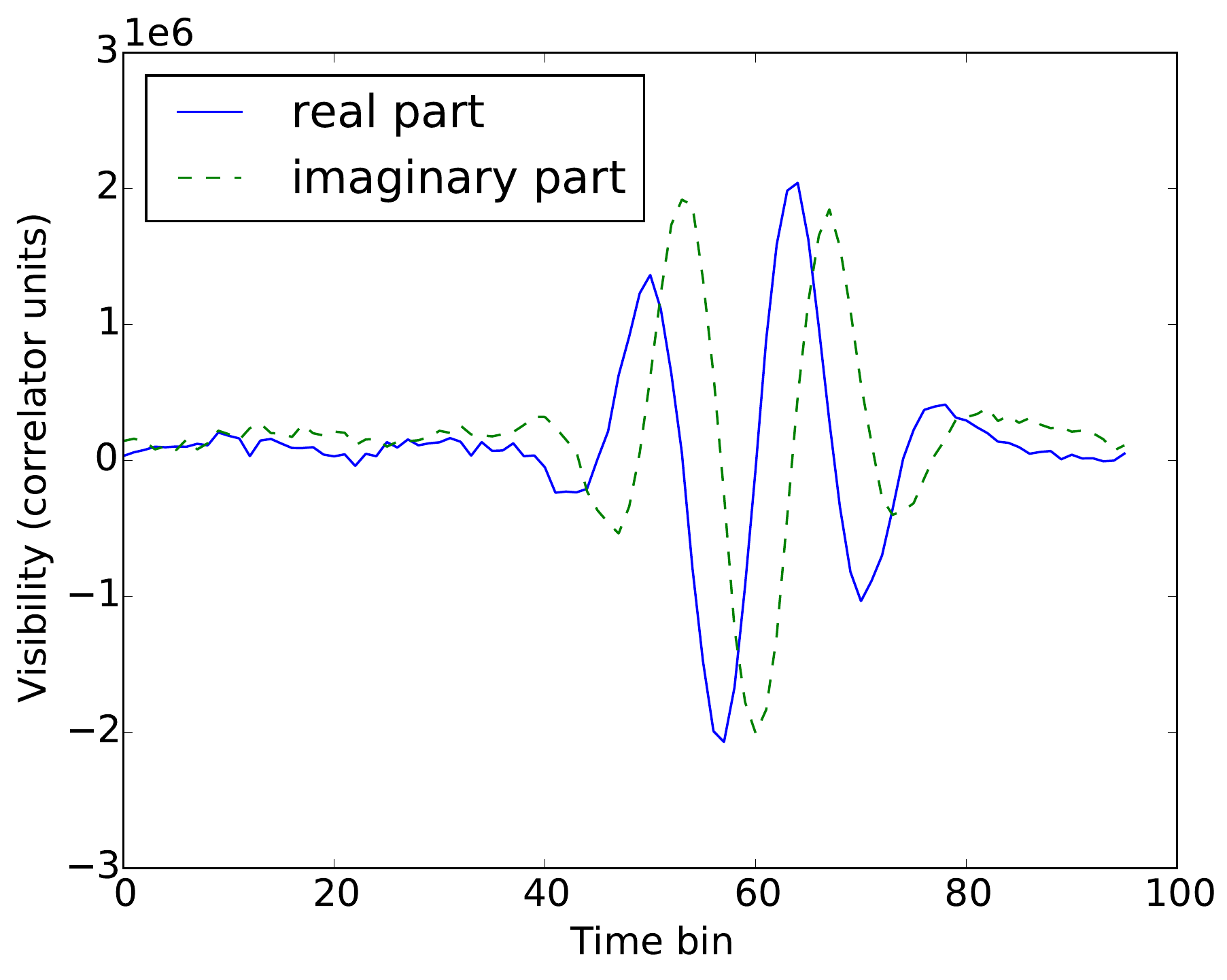}}
    \caption{\label{f:vis}
        Example visibility data for an inter-cylinder baseline of the CHIME
        Pathfinder:
        96 10\,s-integrations of a 0.39\,MHz wide
        spectral bin at
        644\,MHz.
    }
\end{figure}

\begin{figure}
    \centerline{\includegraphics[scale=\bitfigscal]{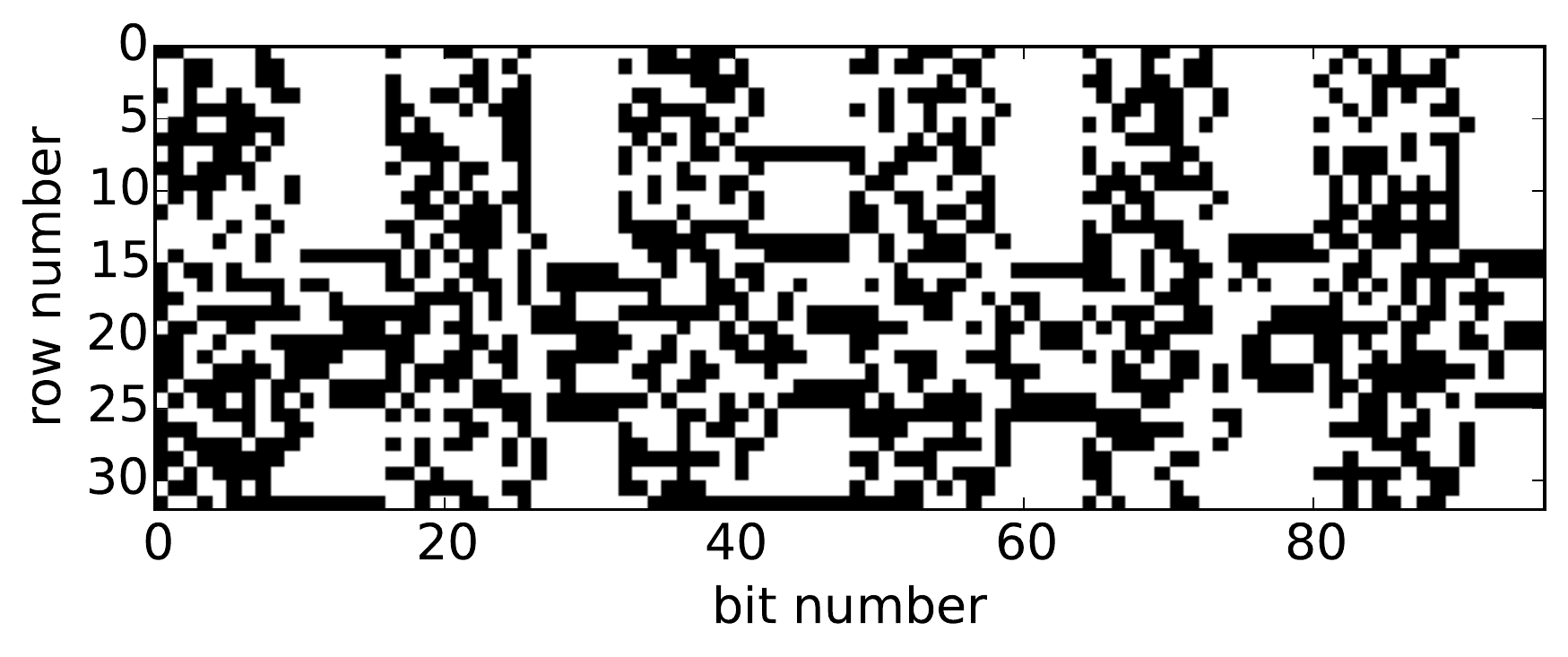}}
    \centerline{\includegraphics[scale=\bitfigscal]{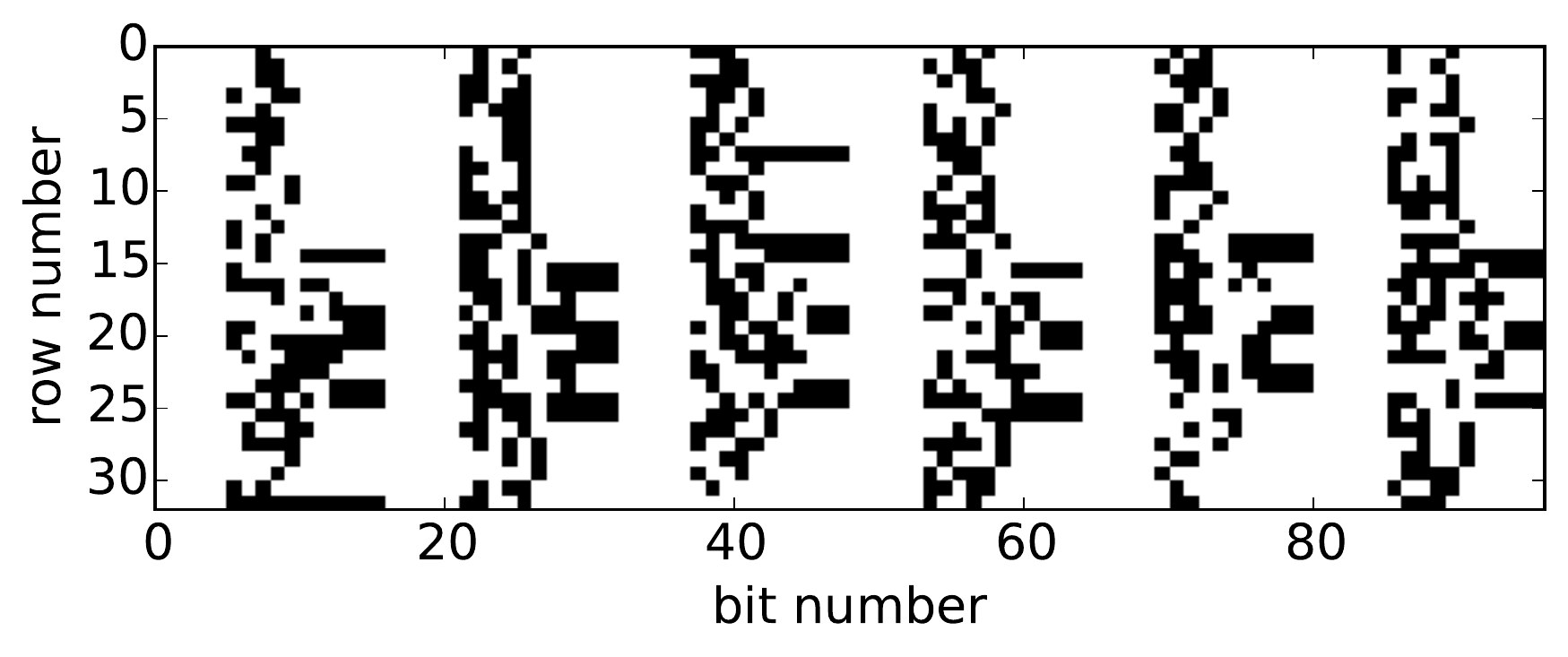}}
    \centerline{\includegraphics[scale=\bitfigscal]{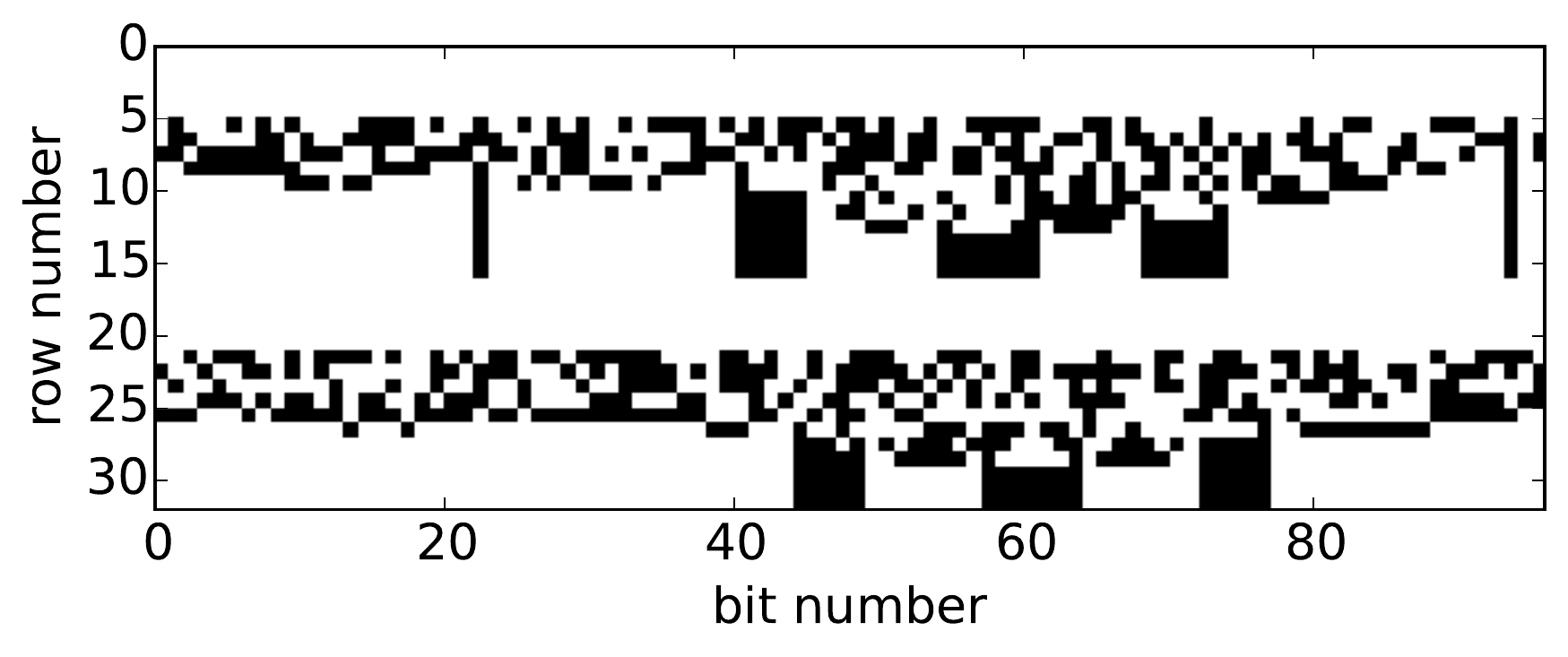}}
    \centerline{\includegraphics[scale=\bitfigscal]{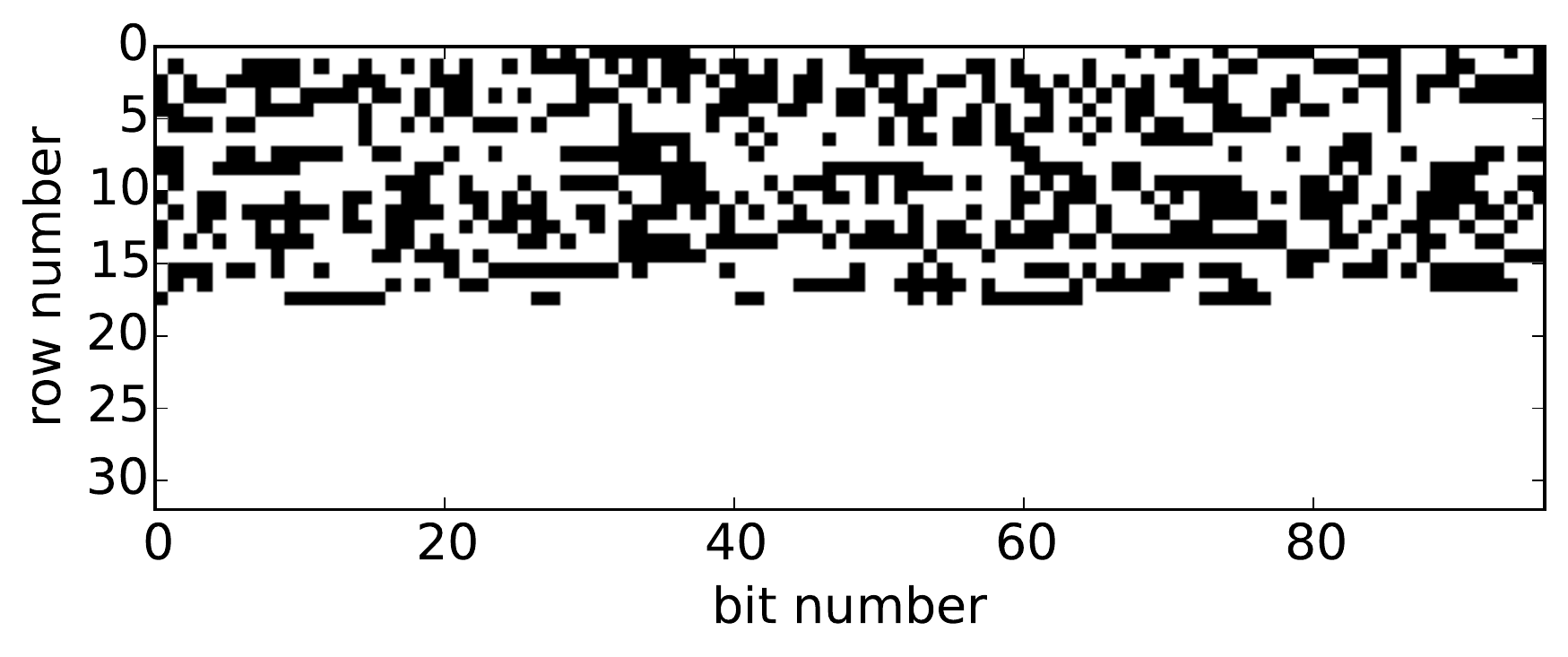}}
    \captionsetup{singlelinecheck=off}
    \caption[bits]{\label{f:bits}
        Bit representation of the data in Figure~\ref{f:vis} at various stages
        of compression. The data are stored as an array of two element structs,
        where each the real and imaginary parts are represented by
        little-endian signed integers.
        The data are natively represented by 32-bit integers, but 
        for the compactness of this figure, we divide the
        data by $2^8$ and use 16-bit integers.
        In all panels, the memory is laid out from 
        left to
        right (in rows of 96 bits) then top to bottom, with black representing
        a set bit and white representing an unset bit.  Within an 8-bit byte
        the bits are unpacked from the least significant bit to the most
        significant bit, which is convenient for visualizing little-endian
        data types. The panels represent, from top to bottom:
        \begin{enumerate}
            \item The original data with each row containing 3 integrations.
            \item Data after reducing precision with $f=0.01$.
            \item Data after the bit-transpose step of \bshuf{}. Each column contains a single integration.
            \item Data after the \lzfour{} compression step of \bshuf{}.
        \end{enumerate}
    }
\end{figure}

The precision reduction applied to the data shown in Figure~\ref{f:vis} is
illustrated in the first two panels of Figure~\ref{f:bits}.

\section{Lossless compression: \bshuf{}}
\label{s:lossless}

Here we discuss lossless data compressors in the context of radio astronomical
data. We seek a compressor that is fast enough for high performance applications 
but also  obtains high compression ratios, especially in the context of the
precision reduction discussed in the previous section.
Satisfying both criteria is difficult and existing compressors are
found to be inadequate. Therefore, a custom compression algorithm, \bshuf{},
was developed; it is both
fast and obtains high compression ratios, at the expense of being slightly less
general.

In this section we begin by reviewing popular algorithms, some
understanding of which is necessary to motivate the design of \bshuf{}. We
then describe the \bshuf{} algorithm, its interaction with the precision
reduction step, and its implementation.

\subsection{Brief description of popular lossless compression algorithms}

\label{s:lossless_review}

By far, the most common class of compression algorithms is the \lzss{} class of
encoders \citep{Ziv1977, Ziv1978}.
These include \lzss{},
\lzf{}\footnote{\url{http://oldhome.schmorp.de/marc/liblzf.html}},
{\tt LZO}\footnote{\url{http://www.oberhumer.com/opensource/lzo/}},
Google's {\tt Snappy}\footnote{\url{https://code.google.com/p/snappy/}},
\lzfour{}\footnote{\url{https://code.google.com/p/lz4/}}
and others.
The \lzss{} encoders compress data by searching for repeated
sequences of bytes in the uncompressed data stream.  When a
sequence that occurred earlier in the stream is found it is
replaced by a reference to the
earlier occurrence. This is represented by a pair of tokens representing
the length of the sequence and the location of the previous occurrence as an
offset from the present location.  It is
worth noting that run-length encoding, where consecutive repetitions
of identical sequences of bytes are eliminated, is a special case and can be
represented by setting the length to be greater than the offset.  This class of
encoders has the advantage that it can be made very fast, with some
implementations (\emph{e.g.}~\lzfour{}) achieving greater than 2\,GB/s 
decompression speed on a single
core of a modern processor.

The \deflate{} algorithm \citep{rfc1951}---best known for its use in both the 
{\tt gzip} and 
{\tt zip} programs and file formats---includes \lzss{} encoding as a first 
step, followed
by Huffman coding \citep{Huffman1952}. Huffman coding entails replacing the most
commonly occurring byte-values in the uncompressed stream with shorter
representations (less than one byte). Less commonly
occurring values must be represented using a symbol that is larger than a byte.
The \lzss{} encoding and Huffman coding are synergetic as they exploit
different types of redundancy. Spatial redundancy
plays a greater role in
\lzss{} as the sequence must be an exact match to a previous occurrence, but
can be compressed even if the bytes-value within the sequence are rare.  The
Huffman coding step exploits only the fact that some bytes may be more common
than others and compresses data even if these bytes are randomly ordered.
Due to the additional Huffman coding step, \deflate{} generally achieves higher
compression ratios than the pure \lzss{} class encoders. However the
computational cost is high and, as we will show,
\deflate{} implementations are generally roughly a
factor of ten slower than the fastest \lzss{} encoders
for both compression
and decompression.

The above algorithms are representative of those most commonly used for
scientific data.  Notable omissions are {\tt
bzip2}\footnote{\url{http://www.bzip.org/}} and {\tt
LZMA}\footnote{\url{http://www.7-zip.org/sdk.html}}, both of
which generally achieve higher compression ratios than \deflate{} but were deemed
too computationally expensive for high-performance applications.

For \emph{typed} binary data, where the data consist of arrays of elements of
a fixed number of bytes, it has been recognized that compression is generally
improved by applying the byte reordering \shuf{} pre-filter \citep{hdf5shuf}.
\shuf{} breaks apart the bytes of each data element, grouping together all the
first (second, etc.) bytes.  To put this in other terms, if you arrange all the
bytes in the array into a matrix with dimensions of the number of elements by
the size of each element, \shuf{} performs a transpose on this matrix.  This
improves compression ratios, primarily in the \lzss{} step, by creating long
runs of highly correlated bytes. This relies on consecutive values of the data
themselves being highly correlated, but this is broadly the case in scientific
data. An illustrative special case is in unsigned integer data that only spans
a subset of the representable values. Any unexercised most-significant-bytes
are grouped together into a long run of zeros and are trivially compressed.

When paired with the precision reduction described in Section~\ref{s:lossy},
it is expected that it is the Huffman coding that will best exploit the
associated reduction of entropy to achieve a better compression
ratio. Even when paired with \shuf{}, the precision reduction does not
generically produce long runs of repeated bytes unless eight or more bits are
discarded.  However, the frequency of certain byte values (multiples of
$2^{n_{BT}}$ where $n_{BT}$ is the number of bits discarded) is greatly
increased; this is a prime target for Huffman coding.

\subsection{\bshuf{} pre-filter and compressor}

\bshuf{} extends the concept of \shuf{} to the bit level: it arranges
the \emph{bits} of a typed data array into a matrix with dimensions of the
number of elements by the size of each element (in bits), then performs a
transpose. This is illustrated in Panel~3 of Figure~\ref{f:bits}.

\bshuf{} is better able to convert spatial correlations into
run-lengths than \shuf{} because it is able to treat correlations within a
subset of the bits in a byte instead of only those which apply to the whole
byte. It thus allows for the elimination of the computationally expensive
Huffman coding step of \deflate{} in favour of the fastest \lzss{}-class
compressor available (we use \lzfour{}).  The trade-off is that,
because each byte now contains
bits from eight neighbouring data elements, spatial correlations must be eight
times longer to produce run-lengths of useful length. An illustration of
bit-transposed data after compression is shown in Panel~4 of Figure~\ref{f:bits}.

While the practice is not widely used, we do not claim to be the first to implement
bit-transposition of data arrays for the purposes of data compression.  In
addition to several references to this idea scattered around the
World Wide Web, the MAFISC compressor \citep{Hbbe2012} implements
bit-transposition as one of its pre-filters.

It is worthwhile to briefly discuss how two's-complement
signed integers are compressed
with \bshuf{}. In two's complement, zero is represented by having none
of the bits in the element set and -1 is represented by having all of them set.
As such, while the values of data having zero crossings may be highly
correlated, the bit representations are not.  Bit-transposing such datasets does
not produce the long runs of zeros or ones in the most significant bits. This
can be clearly seen in Panel~3 of Figure~\ref{f:bits}.
As such, it might be expected that
such data would not compress well.  However, while the sequence of bytes
representing the data's most significant bit (bottom row of Panel~3 of
Figure~\ref{f:bits}) may be incompressible, it is identical to the sequence
of bytes representing the second most significant bit (next to bottom row) and
so on.  As such, the block as a whole turns out to be highly compressible.

\subsubsection{Implementation}

The bit-transpose operation presented here is computationally more expensive than
the byte-transpose in \shuf{} by roughly a factor of four depending on
implementation.  However, both costs are negligible relative to \deflate{}.
\bshuf{}
implements the bit-transpose using the vectorized SSE2 (present on x86
processors since 2001) and AVX2 (present on x86 processors since 2013)
instruction sets when available. Using SSE2
instructions, the most computationally expensive part of the bit-transpose can
process 16 bytes of data in 24 instructions \citep{Sandberg2011}.
Using AVX2 this improves to 32
bytes of data in 24 instructions.  In the absence of these instruction sets,
the bit-transpose is performed using an algorithm
that processes 8 bytes in 18 instructions \citep{hackers}.

For performance reasons, it is beneficial to integrate the lossless compressor, 
\lzfour{}, directly into
\bshuf{} rather than applying it as a sequential HDF5 filter.  The
idea is to bit-transpose a small block of data that fits comfortably into the
L1d memory cache and then apply the compressor while it is still in cache. Since
getting the memory contents in and out of the cache can be the bottleneck,
especially
when using multiple threads, this can greatly improve performance. Usually,
compressing data in small blocks is detrimental to compression ratios, since
the maximum look-back distance for repeated sequences is limited.  However,
because compression after the bit transpose is trivial, this was found not
to be the case for \bshuf{}. The default block-size in \bshuf{}
is 4096\,bytes.

\bshuf{} is both internally threaded using {\tt OpenMP} and thread safe, making
no use of global or static variables.  Threading is implemented by distributing
blocks among threads.

\bshuf{} is written in the C programming language, although it has bindings in
Python and is distributed as a Python package. In addition to routines for
processing raw buffers, it includes an HDF5 filter
which is accessible in Python, can be compiled into a C program, or loaded
dynamically using HDF5's dynamically loaded filters (available in HDF5
version 1.8.11 and later).

\section{Evaluation of method}
\label{s:eval}

In this section we apply the compression algorithm described above to data from
the CHIME Pathfinder
to assess the algorithm's performance and to compare it with other compression
schemes. The Pathfinder comprises two parabolic cylinders, each  20\,m wide by
35\,m long, with their axes running in a north-south direction. 64 identical
dual-polarization feeds are located at 0.3\,m intervals along the central
portion of each focal line. 

The data used for the following comparisons was collected on January 25, 2015,
starting at roughly 2:10AM PDT.
\blue{Analogue signals from the CHIME antenna are digitized at 8 bits before being
Fourier transformed into spectral channels \citep[Section~5]{2014SPIE.9145E..22B} and
correlated \cite{2015arXiv150306202D,2015arXiv150306203K,2015arXiv150306189R}.}
We include data from 16 correlator inputs
connected to four dual-polarization
antennas on each of the two cylinders.
The dataset includes 1024 time integrations of 21.45\,s
length, 136
correlation products, and a subsample of 64 of the 1024 spectral frequencies
uniformly spanning the
400--800\,MHz band.  The data is arranged such that time is the fastest varying index
and frequency the slowest and, as such, the C shape of the array is
{\tt (64, 136, 1024)}. Each element is a struct of two 32-bit, signed,
little-endian integers representing the real and imaginary parts of the
visibility. The total size of the dataset is $64 \times 136 \times 1024
\times 8\,\textrm{bytes} = 68\,\textrm{MiB}$.

The data themselves have a rich set of structure, including spectral channels with
either persistent or intermittent RFI, a malfunctioning amplifier in one of
the 16 signal chains (causing high power and noise in that channel),
and the transit of a bright source (the Crab
Nebula) as well as part of the Galactic
plane. Figure~\ref{f:vis} shows a small subset of the dataset including
the transit of the Crab Nebula in an inter-cylinder baseline. These data
broadly represent the phenomena expected to occur in CHIME data, but
statistically will differ significantly from the data produced by the full
Pathfinder. With 256 correlator inputs, the data produced by the full
Pathfinder will be much more heavily dominated by cross-correlations of long
baselines which may have a significant impact on compression.

For all the tests presented below, we use the HDF5 data format and library to
perform the compression and store the data. The chunk shape is chosen to be
{\tt (8, 8, 1024)} which gives a total size of 512\,KiB.

\subsection{Distribution of rounding errors}

\begin{figure}
    \centerline{\includegraphics[scale=0.45]{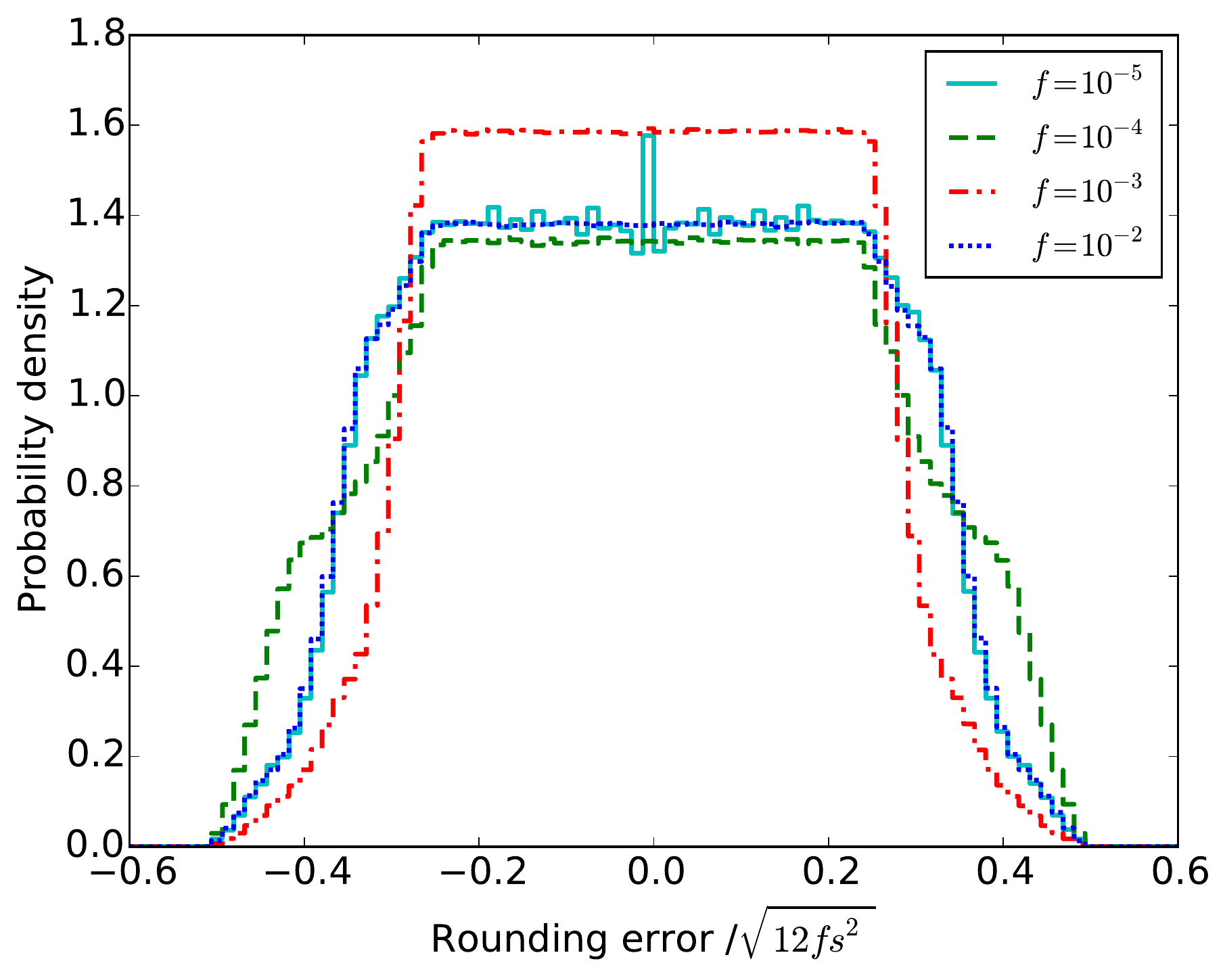}}
    \caption{\label{f:ehist}
        \blue{
        Distribution of rounding errors for different levels of
        precision reduction.  Rounding
        errors are scaled to the maximum granularity, $\sqrt{12fs^2}$, and
        the histogram is normalized to integrate to unity and thus approximates
        the probability density function.
        }
    }
\end{figure}

\blue{
First we verify that our implementation of the precision reduction behaves as
expected when applied to real data.  We require that the rounding errors be
unbiased, and that the probability distribution of errors be more concentrated
than a top-hat function with width of the maximum granularity, given in
Equation~\ref{e:trunc_error}. Rounding errors
are calculated by directly subtracting the original dataset from the precision
reduced dataset then comparing with the maximum granularity. The distribution
of these errors is shown in Figure~\ref{f:ehist} for various values of $f$.
}

\blue{
The expected probability density is a superposition of top hat functions with
widths depending on where the maximum granularity falls relative to a power of
two. The function is flat between $\pm 1/4$ since the final power-of-two 
granularity is always at least half the maximum granularity.
As expected, we see that no rounding error exceeds half the maximum
granularity in absolute value.
For $f=10^{-5}$ there is an excess of errors at zero as well as a
noted jaggedness along the central plateau. This is because, for a significant
portion of the data, the granularity is unity, implying no rounding for
integers. One might notice that the probability densities for $f=10^{-5}$ and
$f=10^{-3}$ are nearly identical. This is because these values of $f$ differ
by a factor very close to a power of four ($4^5$) and a scaling relation 
guarantees identical probability distributions for this case.
}

\blue{
The requirement that the rounding be unbiased is satisfied if the probability
densities are symmetric about zero, which we have verified is true to within
statistical uncertainty from the finite sample. We have also checked for bias
in the mean of the errors over time in each frequency and correlation product,
as well as searched for correlations in the errors along the time axis, finding
no evidence for either.
}

\blue{
We conclude that there is no evidence that the precision reduction is behaving
other than expected. Our tests are consistent with it increasing the
noise by a fraction of at most $f$, equivalent to a fractional loss of 
integration time of $f$.
}

\subsection{Effectively compressing precision-reduced data}

Here, we assess the effectiveness of the precision reduction step and evaluate
which subsequent lossless compression algorithms are able to exploit the
associated reduction in entropy. In this
section, we consider three classes of lossless compression algorithm: the
\lzss{} class of encoders represented by \lzf{} (chosen due to the
availability of an HDF5 filter), the \deflate{}
algorithm (implemented in {\tt zlib}\footnote{\url{http://www.zlib.net/}}
with compression level 4), and \bshuf{}.
For the \lzf{} and \deflate{} cases, the \shuf{}
pre-filter was applied to the data which was found to improve
compression ratios in all cases. We quote compression ratios as the ratio of
compressed data size to original data size, expressed as a percentage.
Results are shown in Table~\ref{t:trunc}.

\begin{table}
    \caption{
        \label{t:trunc}
        Compression ratios for various compression algorithms as a
        function of degree of precision reduction (parameterized by $f$,
        defined in Section~\ref{s:lossy}).
    }
    \begin{tabular}{|c|p{1.9cm} p{1.9cm} p{1.9cm}|}
        \hline
        $f$ & \lzf{} & \deflate{} & \bshuf{} \\
        \hline

$       0$ & 69.5\% & 61.2\% & 59.6\% \\
$ 10^{-5}$ & 46.7\% & 38.5\% & 37.1\% \\
$ 10^{-4}$ & 45.6\% & 34.1\% & 32.0\% \\
$ 10^{-3}$ & 44.2\% & 30.6\% & 27.9\% \\
$ 10^{-2}$ & 37.1\% & 25.9\% & 22.2\% \\

        \hline
    \end{tabular}
\end{table}

It is seen that reducing the precision of the data does in fact improve
compression in all cases. Between $f=10^{-5}$ and $f=10^{-2}$, the precision of
the data is decreased by a factor of $\sqrt{1000}$, or 5.0 bits. Since the
data are stored as 32-bit integers, one would optimistically hope for
a 15.5\% improvement in compression ratio between these cases. All three
compressors do a reasonable job of exploiting the reduced entropy, with \lzf{},
\deflate{}, and \bshuf{}, achieving a 9.6\%, 12.6\%, and 14.9\% increase in
compression respectively. \lzf{}'s marginally poorer ability to exploit the
reduced entropy is in line with our expectations from
Section~\ref{s:lossless_review}.

It can be seen that the compression ratio improvements are more uniform for
\bshuf{} and \deflate{} than for \lzf{}. The former compress by an extra
$\sim4.5\%$ at each step, while \lzf{} sees a much better
improvement between $f=10^{-3}$ and $f=10^{-2}$.  We speculate that this is
because \bshuf{} and \deflate{} effectively compress each bit as it is
discarded, while \lzf{} achieves most of its improvement when the rounding passes a
byte-boundary for some portion of the data.

The improvements in compression from the native precision
case depend on how much precision is kept by the correlator and data collection
software. The CHIME correlator truncates to 4 bits of precision after spectral
channelization, with the rest of the correlation process being very nearly
lossless. We see that this process keeps an excessive amount of precision, since
compression ratios improve by more than 22\% even when reducing the precision
to a conservative $f=10^{-5}$.  While the precision of the data
could in principle be reduced explicitly during acquisition by
right-bit-shifting the values by
several places, this is much less controlled than
comparing to the radiometer equation in the way presented here.

\blue{
It is worthwhile considering what compression ratio would be achieved
if we were to simply reduce the
precision then use a minimum data element size. Such a scheme could be
conveniently implemented using a custom HDF5 data type along with the N-bit
filter. The number of required bits per element is given by
$-\log(12f/N) / (2 \log 2)$ which, for this data and for $f=10^{-3}$, is roughly
15 bits.  One more bit is required for the sign, and at least one more should
be allowed for dynamic range (meaning the total power could change by a factor
of two without overflowing), and so such a scheme would achieve a compression
ratio just worse than 50\%. We see that a mild amount of precision reduction
coupled with lossless compression beats this and does not have the complications
of tuning scaling factors nor worries about overflows during transits of bright
sources or bursts of RFI. In such a scheme changing $N$, which depends on the
integration time and spectral channel bandwidth, would require a change in
data type which is inconvenient.
}

\blue{
In Section~\ref{s:req} we argued that ordering the data with time as the
fastest varying index is beneficial for compression. To verify this, we
repeated the tests in this section using a chunk shape of {\tt (64, 32, 32)}.
As expected, all compression ratios worsen. \bshuf{} is most sensitive to data
ordering with compression ratios worsening by
6\% (for $f=10^{-2}$) to 11\% (for $f=0$), compared to 3\% to 4\%
for the other compressors.
}

The benefits of the precision reduction are substantial. When compressing with
\bshuf{}, reducing the precision to $f=10^{-3}$ more than halves the final data
volume.

\subsection{Throughput and compression ratios}

Here we directly compare several lossless compression schemes on the basis of
compression ratio and throughput, fixing the lossy compression at $f=10^{-3}$.
In addition to the \deflate{}+\shuf{} and
\lzf{}+\shuf{} schemes presented in the previous section, we compare \bshuf{} to
\blosc{}\footnote{\url{http://www.blosc.org/}}.
\blosc{} is actually a `meta-compressor' which combines
an optimized version of the \shuf{} pre-filter with a lossless compressor,
using a similar blocking scheme as \bshuf{} to optimize
the use of the L1d cache. \blosc{} supports several back-end lossless
compressors.  Here we show results for \lzfour{} as well as \lzfourhc{}.
\lzfourhc{} is an \lzfour{} derivative that uses the same compression format
and decompresser as \lzfour{} but spends much longer on the compression step in
attempt to achieve better compression ratios.

Both \deflate{} and \blosc{} have a compression level
parameter, whose value can be between 1 (fastest) and 9 (best compression)
inclusive. For \deflate{} we test levels 1 and 7 to bracket the range of
throughputs and compression ratios that can be expected. Levels higher than 7
were found to be excessively slow at compression while not significantly
improving compression ratios. For the same reasons we
test level 1 for \blosc{}+\lzfour{} and level 5 for \blosc{}+\lzfourhc{}.

To give an idea of the overhead associated with reading and writing to
the datasets, we show the throughput when not compressing the data using both
chunked and contiguous storage. \blue{Finally, to give an idea of the computational
cost of the precision reduction, we include throughputs for the example
implementation, although we reiterate the little effort has been put into its
optimization.}

Benchmarks are for a single thread on an Intel Core i7-3770 CPU running at
3.40GHz. Note that this processor includes support for the SSE2 instruction set
but not AVX2. We employ the HDF5 ``core'' file driver, such that the datasets
only ever exist in memory and are never written to disk.  Thus the file system
plays no role. Our timings are reproducible in repeated trails to within a
few percent.
Results are presented in Table~\ref{t:performance}.

\begin{table}
    \caption{
        \label{t:performance} Comparison of the of various
        compression algorithms on the basis of compression ratios and
        throughput.
    }
    \begin{tabular}{|l r r r |}
        \hline
        \multicolumn{1}{|l}{Algorithm} &
        \multicolumn{1}{c}{Compression} &
        \multicolumn{1}{c}{Write} &
        \multicolumn{1}{c|}{Read} \\
        \multicolumn{1}{|l}{} &
        \multicolumn{1}{c}{ratio} &
        \multicolumn{1}{c}{(MiB/s)} &
        \multicolumn{1}{c|}{(MiB/s)} \\
        \hline

           Contiguous & 100.0\% & 5065 & 2576 \\
              Chunked & 100.0\% & 3325 & 2572 \\
             Rounding & 100.0\% & 289 & 2574 \\
          \deflate~-1 & 32.2\% & 73 & 181 \\
          \deflate~-7 & 30.3\% & 23 & 182 \\
                 \lzf & 44.2\% & 181 & 245 \\
       \blosc+\lzfour & 47.3\% & 528 & 1348 \\
     \blosc+\lzfourhc & 41.0\% & 30 & 1417 \\
               \bshuf & 27.9\% & 749 & 1181 \\

        \hline
    \end{tabular}

\end{table}

We see that \bshuf{} obtains a better compression ratio than all other
algorithms tested and that only \blosc{} outperforms \bshuf{} on read
throughput. The margin by which \bshuf{} outperforms the other compressors is
considerable, producing compressed data two thirds the size of the next best
`fast enough' (by our requirements in Section~\ref{s:req}) compressor, \blosc{}.
In addition, \bshuf{} compresses faster than any compressor tested, which, while not
being a design consideration, is a nice bonus. It is not clear why \bshuf{}
compresses
faster than \blosc{}+\lzfour{}, since they use the same back end compressor,
use similar block sizes and the \shuf{} pre-filter is much less computationally
intensive than \bshuf{}'s bit transpose.

For reading, the HDF5 overhead is significant. Both \blosc{} and \bshuf{} are
within a factor of two of achieving the throughput limits from HDF5.  Put
another way, the HDF5 overhead accounts for more than half the total time
required to read data with these compressors. However, these speeds are all
fast compared to hard-drive read-throughputs.

\section{Summary and conclusions}

We have presented a high-throughput data compression scheme for astronomical radio
data that obtains a very high compression ratio. Our
scheme includes two parts: reducing the precision
of the data in a controlled manner to
discard noisy bits, hence reducing the entropy of the data; and the lossless
compression of the data using the \bshuf{} algorithm.

The entire compression algorithm consists of the following steps, starting with
the precision reduction:
\begin{enumerate}
    \item Estimate the thermal noise on a data-element by data-element
        basis using Equation~\ref{e:noise} in conjunction with the
        noise covariance matrix defined in either
        Equation~\ref{e:rad_simple} or
        Equations~\ref{e:rad_full1}--\ref{e:rad_full3}
    \item Choose an acceptable fractional increase in noise variance $f$; we
        recommend $10^{-5} > f > 10^{-2}$.
    \item Round all data to a multiple of the largest possible power of two 
        subject to the
        limit on rounding granularity given in Equation~\ref{e:trunc_error}.
\end{enumerate}
Followed by the lossless compression:
\begin{enumerate}
    \setcounter{enumi}{3}
    \item Rearrange the bits within blocks of data by arranging them in a matrix
        with dimensions of the number of elements by the size of each element
        (in bits), then perform a transpose.
    \item Compress with a fast lossless \lzss{} style compressor.
\end{enumerate}
The lossless compression is implemented and distributed as the
\bshuf{} software package which includes an HDF5 filter for the algorithm and
bindings for the C and Python programming languages.

We reiterate that the precision reduction and \bshuf{} lossless compression
steps are independent. They work very well together, however,
we have shown that
several commonly available lossless compression algorithms are able to exploit
the reduction of information entropy associated with precision reduction.  In
addition, we have shown that \bshuf{} performs very well compared to other
lossless compressors with and without the precision reduction.

The algorithms in this paper use integers, since the CHIME experiment
produces and records visibility data in integer
representation. However, most of the aforementioned procedures and
conclusions apply equally well to floating-point
data.
\blue{Reducing the precision of floating point numbers entails adapting the
rounding procedure to act only on the significand, and \bshuf{} works for any
data type with no modifications. It is not expected that
floating point numbers will compress as well as integers under this scheme,
especially for data
with frequent zero crossings. In such data there can be large fluctuations in the
exponent, inhibiting compression. This can especially affect
interferometer data where cross correlations over long baselines are often
zero within noise.}

In addition to our compression scheme we present the following considerations
for developing a data storage format for high performance applications:
\begin{itemize}
    \item Because decompression can be fast compared to reading data from disk,
        compression can improve IO performance.
    \item Compression and HDF5 chunking should not reduce the speed of random
        access data reads for files stored on hard disks since the disk seek
        time is generally longer than the time required to read and decompress
        the data.
    \item Data are generally read more often than they are written, and as such
        read performance should be prioritized over write performance. For this
        reason, a post-acquisition data reordering step can be worthwhile.
    \item Data consumer usage patterns should be the primary consideration when
        deciding a data layout. For CHIME this means having time as the minor
        (fastest varying) axis. This is also beneficial for compression ratios
        when using \bshuf{} since time is the axis over which the data are most
        highly correlated.
\end{itemize}

We have shown that when applying our compression scheme to data produced by the
CHIME experiment, the data are compressed to 28\% of their original size
which in many cases will improve read performance by over a factor of three.
In addition we have shown that \bshuf{}, when applied to CHIME data, 
outperforms all compression algorithms tested in both throughput and
compression ratios.

The benefits of our compression scheme are substantial with essentially
no drawbacks. The CHIME experiment is employing the algorithm in a post-acquisition
data re-ordering and compression step that creates our final archive files.
\blue{As radio datasets continue to grow in size, more instruments will need to
employ compression to keep these datasets manageable.  We expect that our
scheme is broadly applicable to most post-correlation radio data, and that
aspects of it could benefit many current and future instruments should the
change in data format be deemed worthwhile.}

\section*{Acknowledgments}

We are very grateful for the warm reception and skillful
help we have received from the staff of the Dominion Radio
Astrophysical Observatory, operated by the National Research Council Canada.

CHIME is Leading Edge Fund project 31170 funded by the Canada
Foundation for Innovation, the B.C. Knowledge Development Fund, ‘le
Cofinancement gouvernement du Qu\'ebec-FCI, and  the Ontario Research Fund.
K.~Masui is supported by the Canadian Institute for Advanced Research, 
Global Scholars Program.
M.~Deng acknowledges a MITACS
fellowship.
We acknowledge support from
the Natural Sciences and Engineering
Research Council of Canada,
the CIfAR Cosmology and Gravity program,
the Canada Research Chairs program, and the National
Research Council of Canada.
We thank Xilinx and the XUP for their generous donations.

\section*{References}

\bibliography{bitshuffle}

\end{document}